\documentstyle[12pt]{article}

\begin{document}
\begin{titlepage}
\title{\bf Thermal and Mechanical Properties of Pt-Rh Alloys}
\date{ }
\author{{\bf  G. Dereli, 
T. \c{C}a\u{g}{\i}n${}^*$, 
M. Uludo\u{g}an, 
M. Tomak}\\  \\  
{\small Department of Physics, Middle East Technical University,}\\ 
{\small 06531 Ankara, Turkey}\\  
{\small ${}^*$Materials and Process Simulation Center, California 
Institute of Technology,}\\ 
{\small Pasadena, CA 91125 , U.S.A.}}
\maketitle
\begin{abstract}
\noindent 
We utilize the many-body potentials developed by Sutton and Chen\\(1990)
within the context of the tight-binding approach to study the bulk
properties of metals and metal alloys in
molecular dynamics (MD) simulations. In the simulations of 
Pt-Rh alloys
we used the MD algorithms based on an extended Hamiltonian 
formalism from the works of Andersen(1980), Parrinello and Rahman(1980), 
Nos\'{e}(1984),
Hoover(1985) and \c{C}a\u{g}{\i}n(1988). The simulator program  
that we use generates information about various
physical properties during the run time, along with
critical trajectory and stepwise information
which need to be analysed post production. The thermodynamical and
mechanical properties are calculated using the statistical
fluctuation expressions over the MD. 
 \end{abstract}
\end{titlepage}
%\newpage
%\renewcommand{\baselinestretch}{1.5}
\noindent {\large {\bf 1. Introduction}} 
\vskip 4mm
The development of advanced high-performance materials in the industrial world
is increasingly coupled with theoretical and computational modelling.
In this process, focus is on research areas 
having direct impact on innovative development of such materials. High-performance 
metallic alloys find use in various segments of materials and 
chemical industries as catalysts, and as low-weight and high-strength structural 
materials.  
In particular, Pt-Rh alloy is important for catalytic reactions
in controlling exhaust gas emissions and NH$_3$ oxidation reactions in fertilizer industries.

The theory and computational efforts require and strive for
i) {\it {a priori}} determination of the ultimate properties of metals, 
metallic alloys, ii)
simulation and modelling of the processing conditions, and 
iii) investigating the performance characteristics of these metals 
and alloys.
All these are extremely important for timely, cost-efficient and 
environmentally 
compliant development of such advanced materials. With advances in 
computational speed and emerging new computational algorithms, 
theoretical 
and computer simulations are positioned in the 
midst of this innovative process.

Computer simulations on various model systems usually use simple
pair potentials. To account for the directionality
of bonding, three-body interactions are also often employed. However, the 
interactions in metals and metal alloys cannot be represented by
simple pairwise interactions.  In these systems the electron density 
plays a dominant role in the interactions and resulting physical properties. 
Therefore interactions in metals and metal alloys are dominated by
many-body interactions.  In simple sp-bonded metals this effect 
may be represented by the interaction potentials derived from model 
pseudopotentials using second-order perturbation theory (Harrison 1979,1980). 
Along these lines, we have developed interaction potentials and utilized 
them to study the properties of simple alkali metal and alkali-metal 
alloys (Dalg{\i}\c{c} et al 1994). However, for d-band metal and metal alloys, the  
model pseudopotential approach gives way to newer techniques evolved 
over the past ten years to account for the many-body effects.  Among these
approaches we can list the empirical many-body potentials based on 
Norskov's Effective Medium Theory (Norskov 1982), Daw and Baskes' Embedded Atom 
Method (Daw and Baskes 1984), Finnis and Sinclair's empirical many-body potentials
(Finnis and Sinclair 1984), and more 
recently the many-body potentials developed by Sutton and Chen(1990)  
within the context of a tight binding approach (Koleske and Sibener 1993).

In this work we  utilized the Sutton-Chen potentials to study 
the bulk properties of Pt-Rh alloy using molecular dynamics 
simulations. 
\vskip 6mm
\noindent {\large  {\bf 2. Many-Body Potentials for fcc Metals and Methods}}
\vskip 4mm

The Sutton-Chen interaction potential is given as the sum 
of a pairwise repulsion term and a many-body density-dependent 
cohesion term (Sutton and Chen 1990).  The functional form of the interaction potential 
is as follows: 
$$U_i = D ({1\over 2}  \sum_j u({\bf r_{ij}})  - c \rho_i ) $$
where
$$u (r) = ({a \over r} )^n  $$
$$\rho_i = (\sum_j \phi(r_{ij}) )^{1\over 2} $$
$$ \phi(r) = ({a \over r} )^m  $$

\noindent The Sutton-Chen potential parameters $D$, $c$ and $m$ and $n$ 
are optimized to fit to the $0$K properties such as 
the cohesive energy, zero-pressure condition and the bulk modulus 
of the fcc metals.  We list the values of 
these parameters for Rh and Pt in Table 1.

As mentioned above, the parametrization of the interaction 
potentials for these fcc metals are based on the 
bulk properties at $0$K.  The functional form is fairly simple
in comparison to Embedded Atom Method potentials and is moderately
long ranged.  The last property makes this set especially
attractive for surface and interface studies amongst others,
since most of them are very short ranged, (i.e. are fitted up to 
first or second nearest-neighbour distances).

The above interaction potential can be generalized to describe 
binary metal alloys in such a way that all the parameters in the
Hamiltonian equations are obtained from the parameters
of pure metals. The Sutton-Chen interaction potential above
is adopted by Rafii-Tabar and Sutton (1991)
to a random fcc alloy model in which sites are occupied by  two types of atoms
completely randomly, such that the alloy has the required average concentration.
The equilibrium lattice parameter $a^*$ at $0$K of the random alloy
is chosen as the universal length scale and the expectation value $E^t$
per atom of the interaction Hamiltonian is given as a function of $a^*$.
Rafii-Tabar and Sutton (1991) determined the value of the equilibrium lattice parameter
for the random alloy, and calculated elastic constants and the enthalpy of 
mixing by the {\it static}-lattice summation method. 
Once $a^*$ is found the enthalpy of mixing $\Delta H$ per atom at $0$K are also 
obtained from
$$\Delta H= E^t - c_A E^A  - c_B E^B$$
where $E^A$ and $E^B$
are the cohesive energies per atom of the elemental $A$ and $B$ metals and the constants are such that $c_A + c_B =1$.

\vskip  4mm
The Molecular Dynamics(MD) and Monte Carlo(MC) simulation methods 
(Allen and Tildesley 1987) have 
long been
very important in studying statistical mechanics of many-body systems.
In recent years, these methods have increasingly been used in studying
more realistic problems to investigate the properties and behaviour
of these systems at elevated temperatures and pressures/applied stresses 
(\c{C}a\u{g}{\i}n 1988a,1988b),
in contrast to lattice dynamics and minimization.  

The algorithmic advances in the late 1980's have enabled researchers to  
simulate the dynamics of open systems for studying the phase equilibria 
(\c{C}a\u{g}{\i}n 1989).  
This is especially important for our goals; for instance the phase segregation 
in metal alloys is one of the key problems in the development of advanced 
high-performance metallic alloys.
\smallskip  

In order to investigate the effect of temperature and concentration on the 
physical properties of random alloys, 
we use molecular dynamics method to calculate the enthalpy of mixing, the density 
and the elastic constants at six different atomic
concentrations $( 0, 20, 40, 60, 80,$ and 
$100 \%$ of Rh in Pt-Rh alloy) 
at temperatures ranging from $T=300$K to $T=1500$K with $200$K increments.  
In contrast to  static lattice sums,
molecular dynamics  inherently takes into account 
the anharmonic effects in computed physical properties.

\smallskip
In the following, we present the expressions specific to these many-body potentials
which are used in the computations.  The many-body force on atom $i$ along a 
direction $a(=x,y,z)$ is given as:
 $$F_{ai}= -{D \over 2} (\sum_j^*  u'(r) {r_{ija}\over r_{ij}} 
 - {c_i\over 2} {\sum_j^* \phi'(r){r_{ija} \over r_{ij} } \over \rho_i} ),$$
where $'$ denotes $ {\partial \over \partial r}$ and $^*$ signifies the exclusion 
of $i=j$ from the sums. 
The anisotropic stress tensor including the contribution from the many-body 
potential is calculated from
$$\Omega P_{ab}= \sum_i { p_{ia} p_{ib} \over m_i }- {D\over 2} 
 \sum_i ( \sum_j^*  u'(r) {r_{ija} r_{ijb}\over r_{ij}} 
 - {c_i\over 2} {\sum_j^* \phi'(r){ r_{ija} r_{ijb}\over r_{ij} } \over \rho_i} ),$$
\bigskip
The potential energy contribution to the elastic 
constants, the hypervirial tensor $\chi_{abcd}$, is given as
\begin{eqnarray}
\Omega \chi_{abcd} &=& {D\over 2} 
  \sum_i ( \sum_j^* ( u'' - {u' \over r_{ij} } ) 
{r_{ija} r_{ijb} r_{ijc} r_{ijd}\over r^2_{ij}} \nonumber \\
& & - {c_i \over 2}  {\sum_j^* (\phi'' - {\phi' \over r_{ij} }) {r_{ija} r_{ijb} 
r_{ijc} r_{ijd}\over r^2_{ij} } \over \rho_i} \nonumber \\
& & + {c_i \over 4} 
{ (\sum_j^* \phi' { r_{ija} r_{ijb}\over r_{ij} } ) 
(\sum_k^* \phi' { r_{ikc} r_{ikd} \over r_{ik} } ) \over \rho_i^3 }).
\nonumber 
\end{eqnarray}
In our computations at each concentration and at each temperature 
we have first  
determined the zero strain state, $h_o$,
of the system by performing constant-temperature
and constant-stress simulations (NPT) at zero stress. This yields the reference
shape and size matrix, $h_o$ in the Parrinello-Rahman (1980) formalism.  
In determining the elastic 
constants, this reference state is used in constant-temperature and constant-volume
simulations (NVE) of 50000 steps for each state point.  
The elastic constants are evaluated using the following statistical fluctuation
formulas (\c{C}a\u{g}{\i}n and Ray 1988)
\begin{eqnarray}
C^T_{abcd} &=& - {\Omega_o \over k_B T } (<P_{ab} P_{cd}> - <P_{ab}><P_{cd}>)
\nonumber \\ & &
+ {2Nk_B T (\delta_{ac} \delta_{bd} + \delta_{ad} \delta_{bc} )\over 
\Omega_o} + <\chi_{abcd}> \nonumber \end{eqnarray}
where $ < > $ denotes the averaging over time and 
$\Omega_o = det h_o $ is the reference volume for the model system.
\smallskip

We use the program Simulator developed by \c Ca\u g\i n  
that employs the state of the art MD algorithms based on
extended Hamiltonian formalisms emerging from the works of
Anderson(1980), Parrinello and Rahman(1980), Nos\'{e}(1984), 
Hoover(1985) and \c{C}a\u{g}{\i}n and Pettitt(1991a,1991b).
We used a 500-atom cubic system and the simulation started 
with atoms randomly
distributed on a fcc lattice. The system is thermalized starting from $1$K
to the target temperature using constant-enthalpy and constant-pressure
(NHP) ensemble by slowly heating while scaling velocities to increment the
temperature of $1$K/step over the specific number of steps depending on the 
target temperature. 
This is followed by strict velocity scaling at each target temperature.
We then performed NPT dynamics at this temperature for 20000 steps
to calculate the volume, density and enthalpy of the system
for each concentration.  The resulting zero-strain averaged matrix 
$<h_0>$  is used in calculating elastic constants over 50000 steps 
of NVE dynamics.
A fifth-order Gear predictor-corrector algorithm is used with $\Delta$t = 2
fs.  The Parrinello-Rahman piston mass parameter
is chosen as W=400 and in NPT runs the Nos\'{e}-Hoover parameter
is set to Q=100.

\vskip 4mm

\noindent {\large {\bf 3. Results and Discussion}}
\vskip 3mm
Here we present molecular simulation results obtained for the Pt-Rh alloy.
We  calculated the density and volume of 500 atoms and 
enthalpy of mixing for Pt-Rh alloy as a function of percent of atomic concentration of
rhodium. We have done the simulation under isothermal-isobaric
condition (NPT) at temperatures ranging from $300$K to $1500$K with
$200$K increments.

In Figure 1, the density and enthalpy at six different atomic concentration 
values  $ 0, 20, 40, 60, 80, 100 \%$ of Rh in Pt-Rh alloy are given.
At the end points we found the density for Rh and Pt to be $12.3 g/{cm^{3}}$ 
and $21.2 g/{cm^{3}}$, respectively, at $300$K. These values show approximately
1.5\% deviation from the experimental values $12.45 g/{cm^{3}}$ and $21.50 g/{cm^{3}}$.
In Figure 2 we have drawn the calculated enthalpy of mixing 
with respect to atomic concentration of Rh in Pt-Rh alloy at two temperature values of
$300$K and $1500$K.

We also studied the thermal expansion behaviour of the Sutton-Chen potential
that we use. The percent change in the lattice parameter at 
each temperature (the lattice parameter at $300$K is used as the reference) 
are given in Table 2.
Comparison with the experiment indicate that for Pt the theoretical
thermal 
expansion coefficient is approximately twice that of the
experimental value  while for Rh it is around 1.3 higher.

In Figure 3 the elastic constants of Pt-Rh alloy
at six different concentration values are shown.
In order to see the effect of temperature, we repeated the drawing at 
three different temperatures ($300$K, $900$K, $1500$K).
Thermal softening of the alloy is observed as the
temperature increases.
In Figure 4  the calculated
bulk modulus values at the same temperatures are shown.

In our calculations at each concentration and at each temperature
we have first determined the zero-strain state of the system by performing 
constant-temperature and constant-stress simulations.
In determining the elastic constants this reference state is used in 
constant-temperature, constant-volume simulations of 50000 steps for each state point.
Comparison with experimental results is possible only at the end points 
where Pt or Rh is pure. This we have done in Table 3. The static results of 
Sutton and Chen(1990) are also given to facilitate comparison.

To summarize, the potentials  used in the present
dynamic simulations take account of the anharmonic effects
and give 
reasonably accurate description of the thermodynamic properties
and the elastic constants of Pt-Rh alloys.
The system is quite stable and the sign of the enthalpy of mixing is correct at all concentration values. For the concentrations and temperatures at which we 
performed our simulations we found
that mixing is enthalpically favourable.
\newpage
\noindent {\large {\bf References}}
\medskip
\begin{description}
\item { } ALLEN,M.P., and TILDESLEY,D.J.,1987, {\bf Computer Simulation of Liquids} 
(Oxford: Oxford U.P.).
\item { } ANDERSON,H.J.,1980, J. Chem. Phys. 72, 2384. 
\item { } \c{C}A\u{G}IN,T.,and  RAY,J.R.,1988, Phys. Rev. B 38, 7940.
\item { } \c{C}A\u{G}IN,T.,1988a,  Phys. Rev. A 37, 199. 
\item { } \c{C}A\u{G}IN,T.,1988b,  Phys. Rev. A 37, 4510. 
\item { } \c{C}A\u{G}IN,T.,1989, "Molecular Dynamics Methods in Studying Phase Equilibria",
in {\bf Computer Aided Innovation of New Materials} Vol.II, p 255-9,  
Eds. M. Doyoma, J. Kihara, M. Tanaka, R. Yamamoto (Amsterdam:North-Holland).
\item { } \c{C}A\u{G}IN,T., and PETTITT,B.M.,1991a, Mol. Phys. 72, 169. 
\item { } \c{C}A\u{G}IN,T., and PETTITT,B.M.,1991a, Molec. Simul. 6, 5. 
\item { } DALGI\c{C},S., DALGI\c{C},S., DEREL\.{I},G.,and  TOMAK,M.,1994, 
Phys. Rev. B, 50, 113. 
\item { } DAW,M.S.,and BASKES,M.L.,1984, Phys. Rev. B, 29, 6443.
\item { } FINNIS,M.W.,and SINCLAIR,J.F.,1984, Phil. Mag. A,50, 45. 
\item { } HARRISON,W.A.,1979, {\bf Solid State Theory} (New York:Dover).
\item { } HARRISON,W.A.,1980, 
{\bf Electronic Structure and Properties of Solids} (New York:Dover).
\item { } HOOVER,W.G., 1985,Phys. Rev. A, 31, 1695.
\item { } KOLESKE,D.D., and SIBENER,S.J.,1993, Surf. Sci. 290, 179.
\item { } NORSKOV,J.K.,1982, Phys. Rev. B, 26, 2875.
\item { } NOS\'{E},S.,1984, J. Chem. Phys., 81, 511.
\item { } PARINELLO,M. and RAHMAN, A.,1980, Phys. Rev. Lett., 45, 1196.
\item { } RAFII-TABAR,H. and SUTTON,A.P.,1991, Phil. Mag. Lett., 63, 217.
\item { } SUTTON,A.P., and CHEN,J.,1990,, Phil. Mag. Lett., 61, 139.

\end{description}
\newpage

\begin{table}[t]
\caption{\label{tab1} The interaction  potential parameters for Pt
and Rh.}
\begin{center}
\begin{tabular}{|c|c|c|c|c|c|}
\hline
&&&&&\\
~~a  &~~~~~~~~~~ $D$~~~~~~~~~~ & c  &~~ m~~ &~~ n~~ &  metal \\
 $(\AA)$ & $(10^{-2}$ eV)&&&& \\
\hline
&&&&&\\
3.92 & 1.98330 & ~34.428 & 8 & 10 & Pt  \\
3.80 & 0.49371 & 145.658 & 6 & 12 & Rh  \\
\hline
\end{tabular}
\end{center}  
\end{table}

\begin{table}[h]
\caption{\label{tab2} The percent change in the lattice parameter
from simulation and experiment. The changes are computed using the
300K lattice parameters from simulation and experiment as
reference, respectively.}
\begin{center}
\begin{tabular}{|c|c|c|c|c|}
\hline
T & \multicolumn{2}{c|} {Pt} & \multicolumn{2}{c|} {Rh} \\ \cline{2-5}
~(K) &  ~This work~ & Experiment &  ~This work~ & Experiment \\
\hline
~500 &  0.35 & 0.18 & 0.26 & 0.18 \\
~700 &  0.73 & 0.38 & 0.54 & 0.39 \\
~750 &  0.83 & 0.42 & 0.61 & 0.46 \\
~900 &  1.14 & 0.59 & 0.83 & 0.61 \\
1000 &  1.36 & 0.70 & 0.98 & 0.73 \\
1100 &  1.59 & 0.81 & 1.13 & 0.84 \\
1300 &  2.10 & 1.04 & 1.46 & 1.11 \\
1500 &  2.72 & 1.29 & 1.80 & 1.38 \\
\hline
\end{tabular}
\end{center}
\end{table}

\begin{table}[t]
\caption{\label{tab3} Elastic constants and bulk modulus of Pt and
Rh. All values are in units of eV A$^{-3}$. At each entry, the
first number gives the molecular dynamics simulation result at
300K. The second number in round brackets is the corresponding
experimental value at 0K, while the third number in square
brackets is the statically calculated value from Sutton and Chen
(1990) at 0K.}
\begin{center}
\begin{tabular}{|l|c|c|c|c|}
\hline
&&&&\\
~~  &~~~~~~~~~~ $C_{11}$~~~~~~~~~~ &$ C_{12}$  &~~$ C_{44}$~~ &~~$ B$~~ \\
\hline 
Pt & 1.81(2.23)[1.96]&1.50(1.59)[1.61]&0.41(0.48)[0.46]&1.60(1.80)[1.73]
\cr
Rh & 2.01(2.63)[2.12]&1.39(1.20)[1.45]&0.82(1.21)[0.89]&1.60(1.68)[1.68]
\cr
\hline
\end{tabular}
\end{center}  
\vspace{5cm}
\end{table}

\newpage

%\noindent {\large {\bf Figures }}

\begin{figure}
\vspace{7cm}
\includegraphics{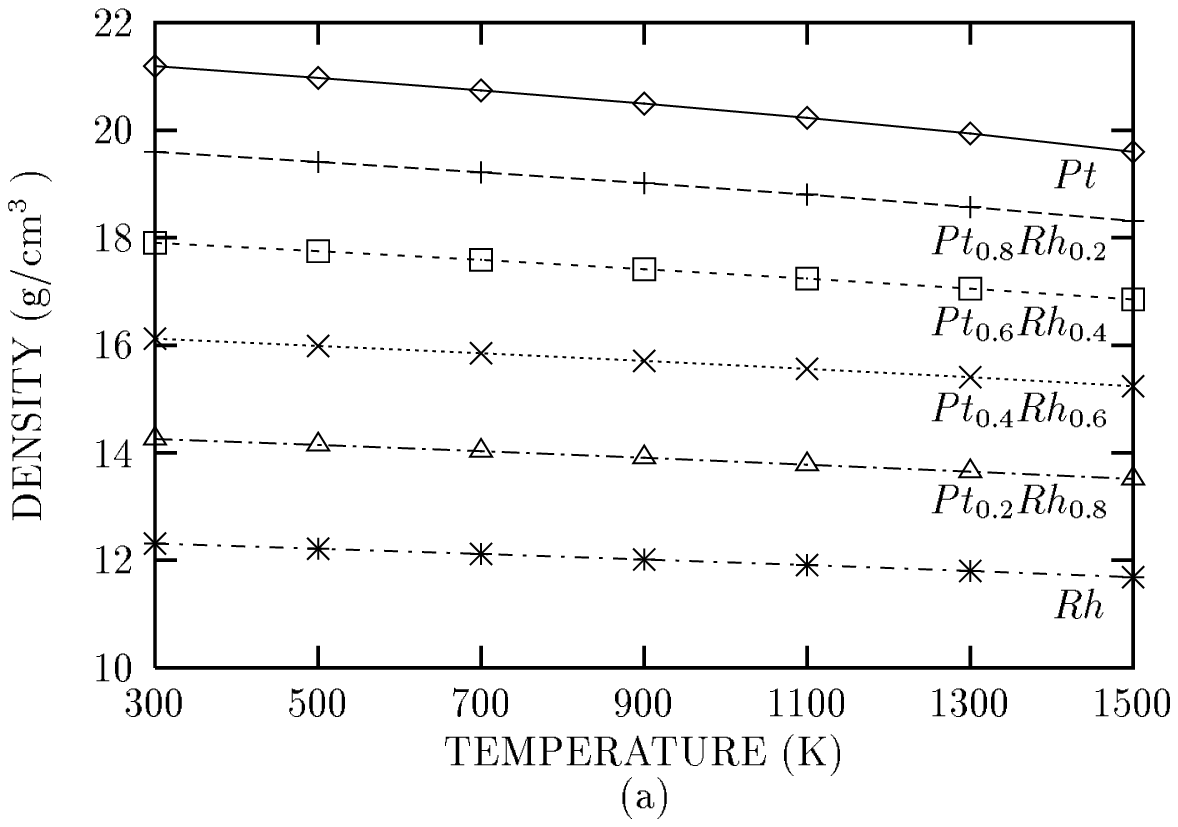}
\vspace{-2cm}
\end{figure}

\begin{figure}
\vspace{7cm}
\includegraphics{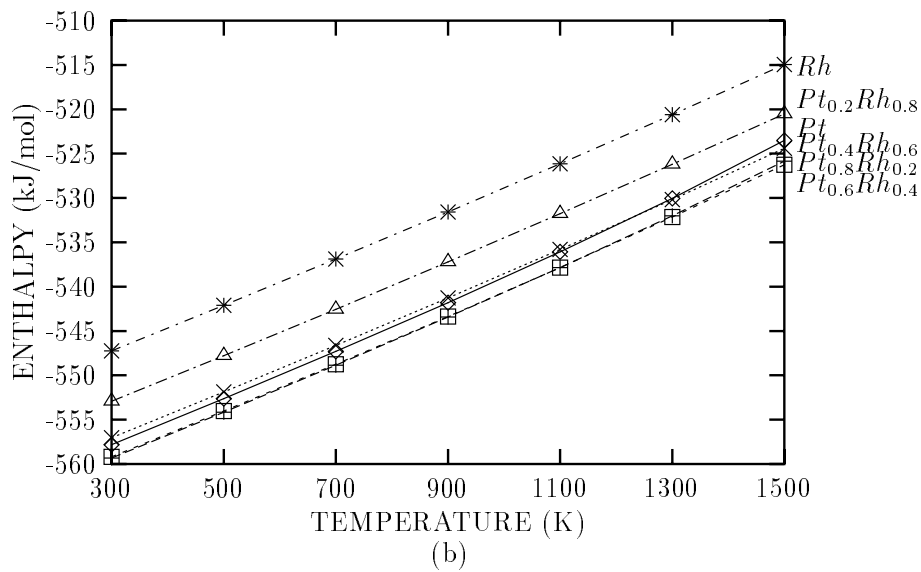}
\vspace{-2cm}
\caption{ (a) Density and (b) enthalpy as a function of temperature
at different atomic concentrations of Rh in  Pt-Rh alloy.
\label{fi1} }
\end{figure}

\begin{figure}[h]
\vspace{7cm}
\includegraphics{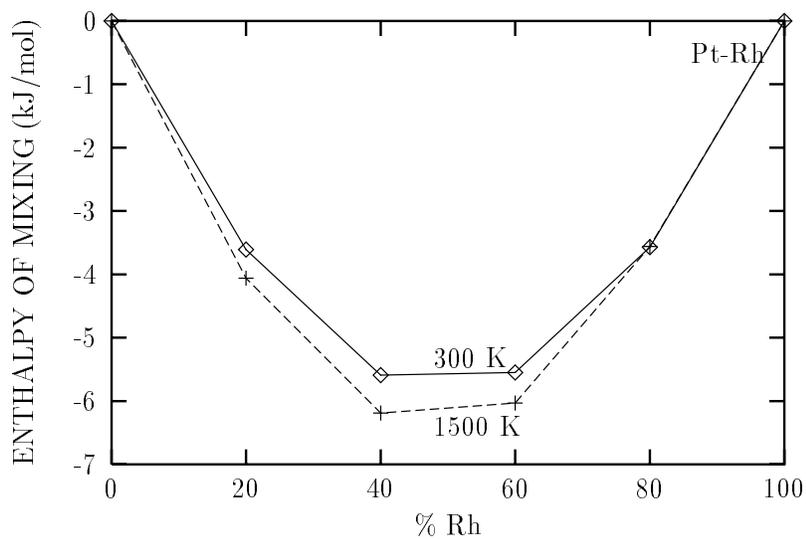}
\vspace{-2cm}
\caption{Enthalpy of mixing as a function of atomic
concentration of Rh in Pt-Rh alloy at two different temperatures.
\label{fi2}}
\end{figure}

\begin{figure}
\vspace{7cm}
\includegraphics{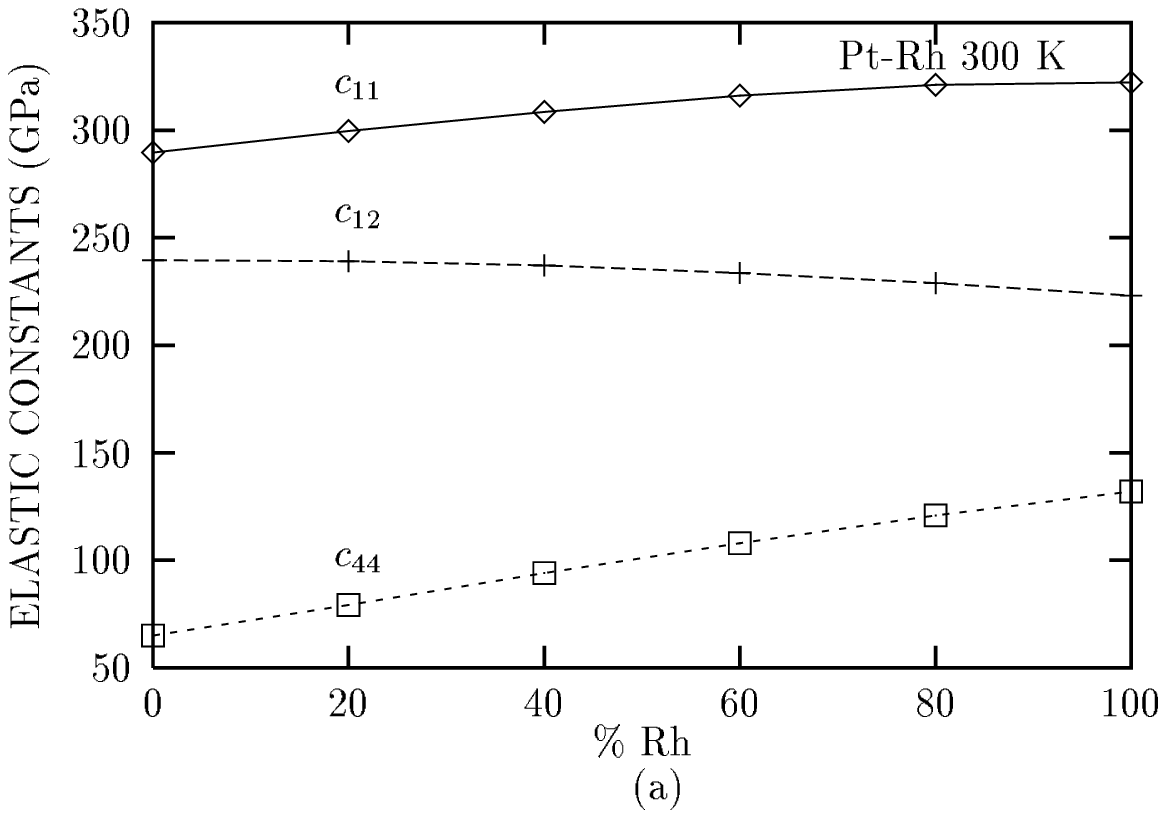}
\end{figure}

\begin{figure}
\vspace{7cm}
\includegraphics{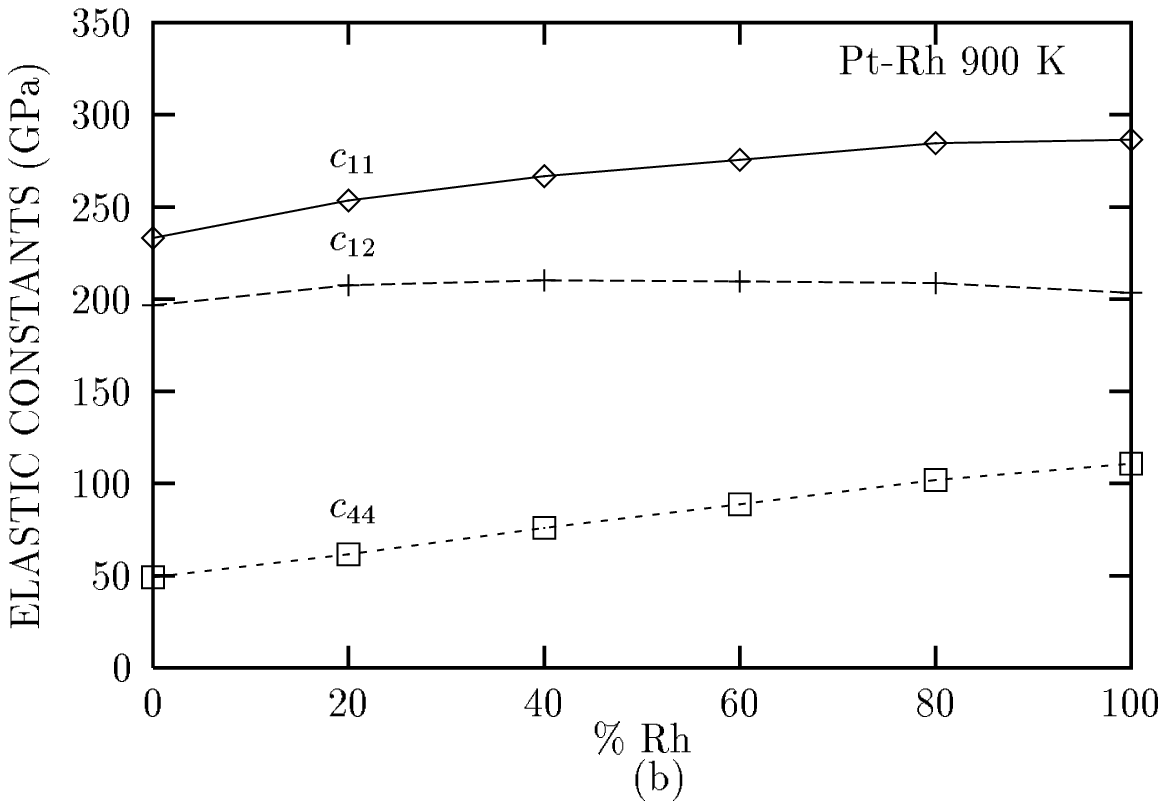}
\end{figure}

\begin{figure}
\vspace{7cm}
\includegraphics{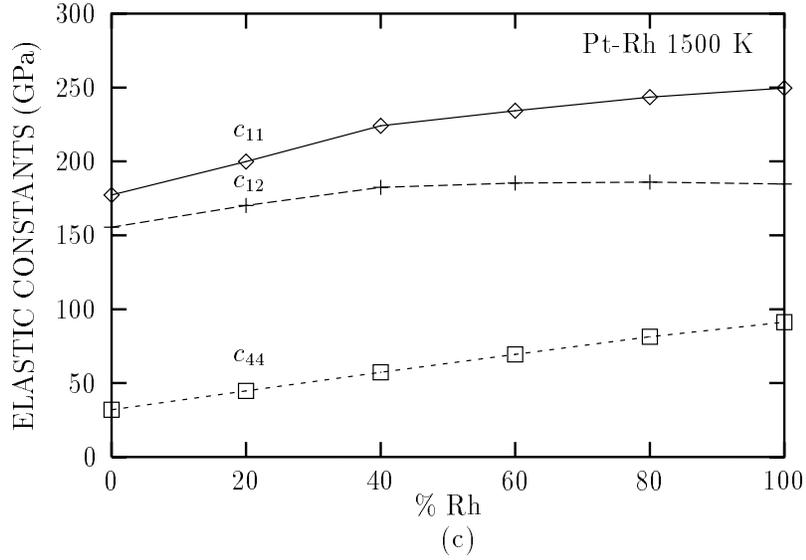}
\vspace{-2cm}
\caption{(a)-(b)-(c) Elastic constants as a function of atomic
concentration
of Rh in Pt-Rh alloy at three different temperatures \label{fi3}}
\end{figure}

\begin{figure}
\vspace{7cm}
\includegraphics{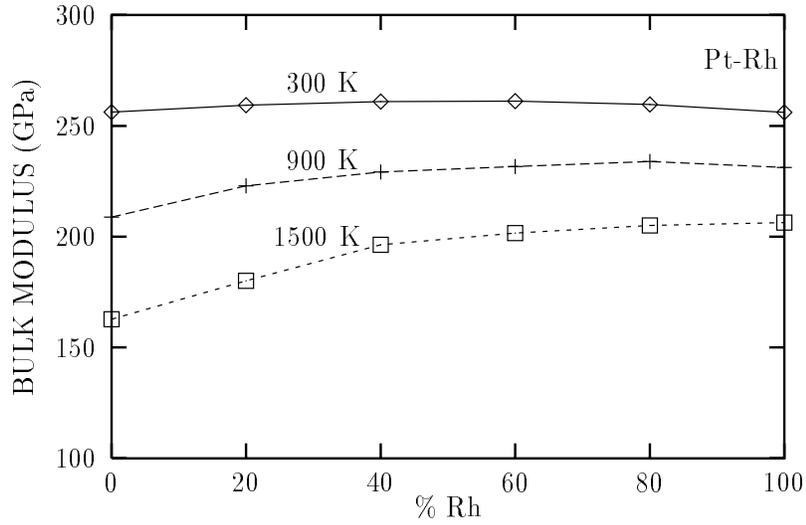}
\vspace{-2cm}
\caption{Bulk modulus as a function of atomic concentration
of Rh in Pt-Rh alloy at three different temperatures.\label{fi4}}
\end{figure}

\end{document}